# Constructing the Critical Curve for a Symmetric Two-Layer Ising Model

M. Ghaemi[1, 2], B. Mirza[3] and G. A. Parsafar[4]

1- Atomic Energy Organization of Iran, Deputy in Nuclear Fuel Production, Tehran, IRAN

2- Chemistry Department, Teacher Training University, Tehran, IRAN
Email: ghaemi@saba.tmu.ac.ir

3- Department of Physics, Isfahan University of Technology, Isfahan 84154, Iran
Email: b.mirza@cc.iut.ac.ir

4-College of Chemistry, Sharif University of Technology, Azadi St. Tehran, Iran
Email: parsafar@sharif.edu

**Abstract**

A numerical method based on the transfer matrix method is developed to calculate the critical temperature of two-layer Ising ferromagnet with a weak inter-layer coupling. The reduced internal energy per site has been accurately calculated for symmetric ferromagnetic case, with the nearest neighbor coupling $K_1 = K_2 = K$ (where $K_1$ and $K_2$ are the nearest neighbor interaction in the first and second layers, respectively) with inter layer coupling $J$. The critical temperature as a function of the inter-layer coupling $\xi = \dfrac{J}{K} \ll 1$, is obtained for very weak inter-layer interactions, $\xi \leq 0.1$. Also a different function is given for the case of the strong inter-layer interactions ($\xi > 1$). The importance of these relations is due to the fact that there is no well tabulated data for the critical points versus $J/K$. We find the value of the shift exponent $\phi = \gamma$ is 1.74 for the system with the same intra-layer interaction and 0.5 for the system with different intra-layer interactions.

**1. Introduction**

The magnetic properties of low dimensional systems have been the subject of intense theoretical and experimental research in recent years. Since exact solution for realistic layer systems on regular lattices are generally unavailable, one relies on approximation schemes to obtain a quantitative picture of the critical data. The two-layer Ising model, as a simple generalization of the 2-D Ising model has of long [1-4] been studied. The two-layer Ising model as a simple model for the magnetic ultra-thin film has various possible applications to real physical materials. For example, it has been found that capping PtCo in TbFeCo to form a two-layer structure has applicable features, for instance, raising the Curie temperature and reducing the switching fields for magneto optical disks [5]. The Cobalt films grown on a Cu (100) crystal have highly anisotropic magnetization [6] and could be viewed as layered Ising models.
From the theoretical viewpoint, the two-layer Ising model as an intermediate between 2-D and 3-D Ising models, is important for the investigation of crossover from the 2-D Ising model to the 3-D Ising model. In particular, it has been argued that the critical point of the latter could be found from the spectrum of the former[7]. In recent years, some approximation methods have been applied to this model[8-11]. It is argued that the symmetric two-layer Ising model is in the same universality class as the 2-D Ising model.[12] Although there are some arguments that the critical exponents may be changed along the critical curve, it has been proposed that the critical exponents would vary continuously along the critical line[12,13]. In a recent work, Ghaemi et al.[14] used the transfer matrix method in the numerical calculation of the critical temperature of Ising and three-state Potts ferromagnet models, on strips of $r$ wide sites of square, honeycomb and triangular lattices in the absence of a magnetic field. They have shown that the existence of the duality relation for the two-dimensional Ising model implies that the reduced internal energy per site, $u(K)$, at the critical temperature must be independent of the size of lattice where $K = j/kT$ and $j > 0$ is the coupling energy for the nearest neighbor spins. Owing to this fact, the calculated $u(K)$ curves plotted versus the reduced temperature, $K$, for the models with various sizes, they were intersected at a single point known as the critical temperature. These authors were able to calculate the critical temperature of the square, triangular, and honeycomb lattices. They then extended the method to the 3-state Potts models to obtain the critical point for the 2-D lattices.[14] In yet another extension, they applied it to the 3-D Ising model.[14] The advantage of this method in addition to its simplicity is that it is only sufficient to know the largest eigenvalue in order to compute the critical temperature. In another work, Ghaemi et al.[15] extended the method given in his previous work[14] to calculate the critical temperature numerically for the anisotropic two-layer Ising ferromagnet, on strip of $r$ wide sites of square lattice. Computation of the largest eigenvalue of the transfer matrix for a semi-infinite two-layer strip of width $r$ becomes difficult when $r$ has large values (namely, larger than 5). In these cases, the size of the transfer matrix is larger than $2^{10} \times 2^{10}$ and hence the computation of the largest eigenvalue ($\lambda_{max}$) is difficult. They have employed the method used by others[16–19] to reduce the transfer matrix size. Rather than diagonalizing the $2^{2r} \times 2^{2r}$ transfer matrix for $r = 3, 4, 5, 6$ and 7, they diagonalized matrices of the rank 8, 22, 44, 135 and 362, respectively. As seen, the size of the reduced transfer matrix is much smaller than that of the original one, such that the $\lambda_{max}$ can be easily calculated from the reduced matrix. Similar to the 3-D Ising models, according to their results, the location of the intersection point depends on the lattice sizes, in other words there is no single intersection point. They have found that the location of the intersection point versus the lattice sizes can be fitted on a power series in terms of the lattice sizes. They have extrapolated the line to an infinite lattice size in order to estimate the critical temperature of the two-layer lattice.

In the present work we want to obtain the critical point as a function of $J/K$ for the symmetric two-layer Ising model, where $J$ and $K$ are the inter-layer and intra-layer interaction energies respectively. The importance of this relation is due to the fact that there is no well tabulated data for the critical points versus $J/K$. On the other hand for any numerical computation using simulation methods we need to know the location of critical point, at instant. Therefore a mathematical relation between the critical points as a function of $J/K$ is very helpful. In the present work we have used the scaling method given in Refs. 16–19 to calculate the critical temperature numerically for the symmetric two-layer Ising model (a special case of anisotropic two-layer model[15]) with a weak inter-layer coupling. In this paper we will construct the critical curve for the symmetric two layer Ising model, it is a smooth curve and we expect that this behavior remains even in the weak decoupling limit. In the asymmetric case there may be discontinuities in the values of the critical exponents which has been mentioned by van Leeuwen.[20]

.

## 2. The Critical Curve for a Symmetric Two-Layer Ising Model

Consider a two-layer square lattice with the periodic boundary condition composed of slices, each with two layers, each layer with $p$ rows, where each row has $r$ sites. Each slice has then $2 \times p \times r = N$ sites and the coordination number of all sites is the same (namely 5). In the two-layer Ising model, for any site we define a spin variable $\sigma^{1(2)}(i,j) = \pm 1$, in such a way that $i=1,..., r$ and $j=1,..., p$, where subscript 1(2) denote the layer number. We include the periodic boundary condition as:

$$\sigma^{1(2)}(i+r, j) = \sigma^{1(2)}(i, j) \tag{1}$$

$$\sigma^{1(2)}(i, j+p) = \sigma^{1(2)}(i, j) \tag{2}$$

In this paper, we discuss the symmetric ferromagnetic case with the nearest neighbor coupling $K_1 = K_2 = K$, where $K_1$ and $K_2$ are the nearest neighbor interaction in the first and second layers, respectively and with inter layer coupling $J$. We take only the interactions among the nearest neighbors into account. The configurational energy for the model may be defined as,

$$\frac{E(\sigma)}{kT} = -K \sum_{i=1}^{r,*} \sum_{j=1}^{p,*} \sum_{n=1}^{2} \{\sigma^n(i,j)\sigma^n(i+1,j) + \sigma^n(i,j)\sigma^n(i,j+1)\}$$

$$- J \sum_{i=1}^{r} \sum_{j=1}^{p} \sigma^1(i,j)\sigma^2(i,j) \tag{3}$$

where * indicates the periodic boundary conditions (eqs.1, 2). The canonical partition function, $Z(K)$, is

$$Z(K) = \sum_{\{\sigma\}} e^{\frac{-E(\sigma)}{kT}} \tag{4}$$

Substitution of eq. 3 into eq. 4 gives,

$$Z(K) = \sum_{\sigma(\{i\},1)} ... \sum_{\sigma(\{i\},p)} \langle 1|B|2\rangle\langle 2|B|3\rangle...\langle p|B|1\rangle \tag{5}$$

where

$$|j\rangle = |\sigma^1(1,j)\rangle \otimes |\sigma^2(1,j)\rangle \otimes ... \otimes |\sigma^2(r,j)\rangle \tag{6}$$

$$\sum_{\sigma(\{i\},j)} = \sum_{\sigma^1(1,j)} \sum_{\sigma^1(2,j)} \cdots \sum_{\sigma^2(r,j)} \qquad (7)$$

By orthogonal transformation, the **B** matrix can be diagonalized, where eq. 4 for the large values of $p$ can be written as

$$Z(K) = \text{tr} B^p \approx (\lambda_{max})^p \qquad (8)$$

where the $\lambda_{max}$ is the largest eigenvalue of **B**. From the well known thermodynamic relation for the Helmholtz free energy, $A = -kT \ln Z$, along with eq. 8 the following results can be obtained:

$$a(K) = -\frac{A}{NkT} = \frac{\ln \lambda_{max}}{r} \qquad (9)$$

$$u(K) = \frac{-E}{Nj} = \frac{\partial a(K)}{\partial K} \qquad (10)$$

where $u(K)$ and $a(K)$ are the reduced internal energy and Helmholtz free energy per site, respectively.

For the two-layer square lattice with size $r$, by using eq. 8, the elements of the **B** matrix have been calculated numerically. We have employed the method of refs. 16-19 for reducing the size of transfer matrix and the $\lambda_{max}$ was calculated with a high precision for different $K$ and $J$ values. The reduced internal energy has been calculated for the two-layer square lattice with ($r = 3, 4,\ldots, 8$) for different values of $\xi$. In order to obtain the value of the critical temperature $K_C$, the intersection point for two unlimited lattices with different sizes should be found.[14, 15] However, such a point may be predicted if we have an expression for the intersection point in terms of $1/n$, where $n = 4rr'$. As shown by in an earlier work,[15] the critical temperature may be given as a general polynomial of degree three as,

$$K_n = \sum_{j=0}^{3} a_j \left(\frac{1}{n}\right)^j \qquad (11)$$

where the value of $a_j$s have been calculated using the least square method.[15] If eq.11 is applicable for large lattice sizes then for the limit of $n \to \infty$, $K_c$ is equal to $a_0$. Such calculations have been done for different values of $\xi \ll 1$. The results are given in the Table 1. As it is seen in Fig. 1, for the weak coupling limit the critical temperature can be written as a simple function of $\xi$.

$$K_C(\xi) = c_0 + c_1\xi + c_2\xi^2 + c_3\xi^3 + c_4\xi^4 + c_5\xi^5 \tag{12}$$

where $C_0 = 0.44069$, $C_1 = -0.46537$, $C_2 = -33.16427$, $C_3 = 1.06375 \times 10^{+3}$, $C_4 = -1.17996 \times 10^{+4}$ and $C_5 = 4.48285 \times 10^{+5}$.

As shown In Figure 1, eq. 12 has a decay form and covers all calculated data for $\xi < 0.1$. Such a calculation had also been performed for different values of $\xi$ larger than 1. The results are given in the Table 2. As it is seen in Figure 2, these points may be fitted with enough precision with the following polynomial:

$$K_c = b_0 + b_1(\xi-1) + b_2(\xi-1)^2 + b_3(\xi-1)^3 + b_4(\xi-1)^4 + b_5(\xi-1)^5 \tag{13}$$

where $b_0 = 0.31211$, $b_1 = -0.05182$, $b_2 = 0.02115$, $b_3 = -5.58026 \times 10^{-3}$, $b_4 = -2.4226 \times 10^{-4}$ and $b_5 = 3.3748 \times 10^{-3}$.

## 2.1. *Shift Exponent*

In this model an interesting situation appears when the inter-layer coupling $J$ becomes in.nitesimally small compared to the inter-layer coupling $K1$ and $K2$. For such cases some scaling theories were constructed.[4,11,13,21,22] They predict the values of the shift exponent $\phi$ which describes the deviation of the critical temperature $Tc(J)$ from the critical temperature in the decoupled limit ($J = 0$)

$$T_c(J) - T_c(0) \sim J^{1/\phi} \tag{14}$$

In particular, these theories predict that when the coupling $K$ is the same in ach layer, then $\phi = \gamma$, where $\gamma$ is the critical exponent describing divergence of susceptibility upon approaching the critical point. We have estimated the shift exponent by using the data in Table 1 and a short program, which is written in Mathematica:

```
<< Statistics_Nonlinear Fit'

Clear[a, b, x];

data = {{0.001, (0.44059)^-1 − ((1/2)Log[1 + 2^{1/2}])^-1},
        {0.003, (0.43975)^-1 − ((1/2)Log[1 + 2^{1/2}])^-1}, . . .}

Nonlinear Fit [data, ax^b, {x}, {a, b}]
```

which has the following result:

$$0.933799\, x^{0.573842},$$

and where the shift exponent is given by

$$0.573842^{-1} = 1.74264,$$

which is in agreement with the arguments that the two-layer Ising model is in the same universality class as the two dimensional Ising model $\phi = \gamma = 1.75$.[23] As there is no exact solution for the two-layer Ising model it is important to calculate the critical exponents by different methods. Our result is deviate from the exact result for the two-dimensional Ising model by 1% . We find $\phi = 0.5004$ for the system with different intra-layer interactions.

## 3. Conclusion

In this paper a numerical method has been used to calculate the critical temperature of the two-layer Ising model which includes the weak coupling limit. The critical points as a function of the inter-layer energies *J/K* have been obtained, the importance of this relation is due to lack of a well tabulated data for the critical points versus *J/K*. Unlike the other approaches given in the literatures, the advantage of this method (in addition to its simplicity) is that only the largest eigenvalue is needed to compute the critical temperature.

We have obtained the critical line as a fifth-order polynomial of $\hat{\iota}$ for the weak coupling limit. The importance of the fifth-order polynomial is due to the fact that, the experimental data could easily be .tted into the polynomial, to obtain a unique value for a physical property. One may extend these calculations to other lattices such as honeycomb, triangular, and also other models (Potts and 3-D Ising models, asymmetric two-layer models, . . .) and use the results for modeling physical systems. It is interesting to have an analytic proof for values of the coefficients $c_i$ and $b_i$ for $i = 1, 2, \ldots, 5$. We hope to make some progress in this direction in near future.

**TABLE CAPTION**

Table1

The Calculated Critical Temperatures of the Two-Layer Ising Model for Different Values of $\xi = \dfrac{J}{K} \ll 1$.

Table2

The Calculated Critical Temperatures of the Two-Layer Ising Model for Different Values of $\xi$ larger than 1.

Table1

| $\xi$ | Calculated critical temperature |
|---|---|
| 0.001 | 0.44059±0.00001 |
| 0.003 | 0.43975±0.00001 |
| 0.005 | 0.43832±0.00001 |
| 0.007 | 0.43638±0.00001 |
| 0.009 | 0.43424±0.00001 |
| 0.01 | 0.43314±0.00001 |
| 0.03 | 0.41669±0.00001 |
| 0.05 | 0.40856±0.00001 |
| 0.07 | 0.40178±0.00001 |
| 0.09 | 0.39664±0.00001 |
| 0.1 | 0.39442±0.00001 |

Table2

| $\xi$ | Calculated critical temperature |
|---|---|
| 1.00 | 0.31211±0.00001 |
| 1.01 | 0.31159±0.00001 |
| 1.02 | 0.31108±0.00001 |
| 1.03 | 0.31059±0.00001 |
| 1.04 | 0.31009±0.00001 |
| 1.05 | 0.30958±0.00001 |
| 1.06 | 0.30908±0.00001 |
| 1.07 | 0.30858±0.00001 |
| 1.1111 | 0.30659±0.00001 |
| 1.1765 | 0.30361±0.00001 |
| 1.2500 | 0.30038±0.00001 |
| 1.3333 | 0.29699±0.00001 |
| 1.4286 | 0.29334±0.00001 |
| 2.0000 | 0.27595±0.00001 |
| 3.3333 | 0.25161±0.00001 |

**FIGURE CAPTIONS**

Figure. 1

The $K_c$ values versus $\xi$, for $\xi \ll 1$ which is fitted into a polynomial with degree of five (doted line).

Figure. 2

The $K_c$ values versus $\xi - 1$, for $\xi$ larger than one which is fitted into a polynomial with degree of five (doted line).

Figure 1

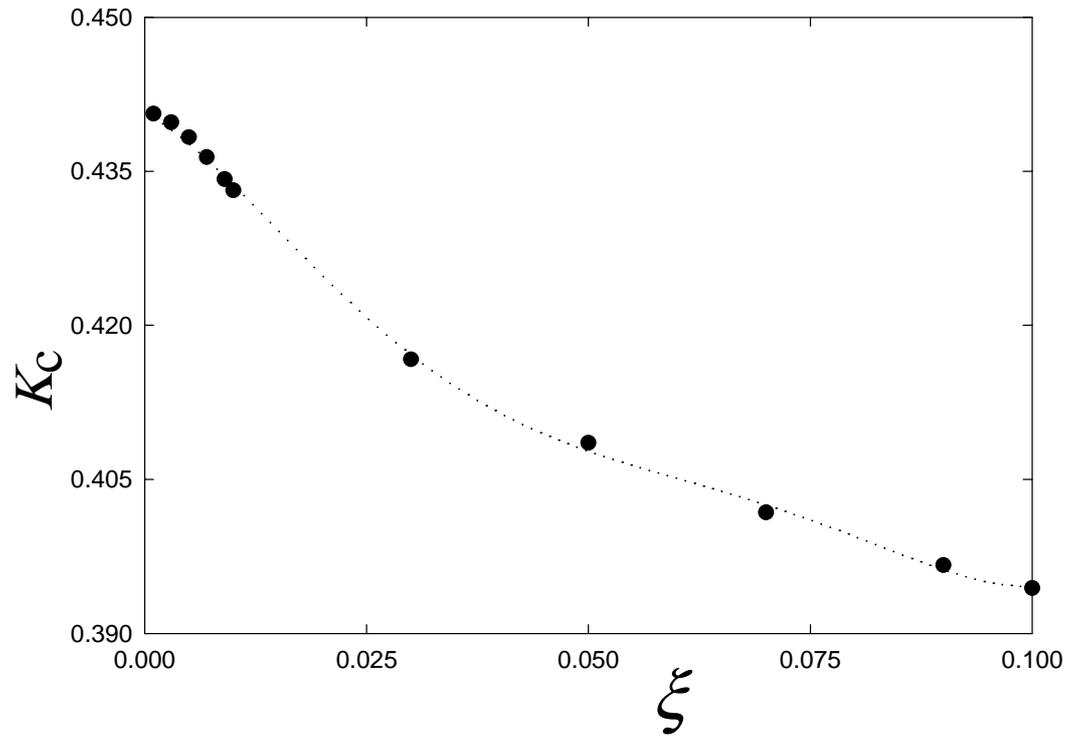

Figure 2

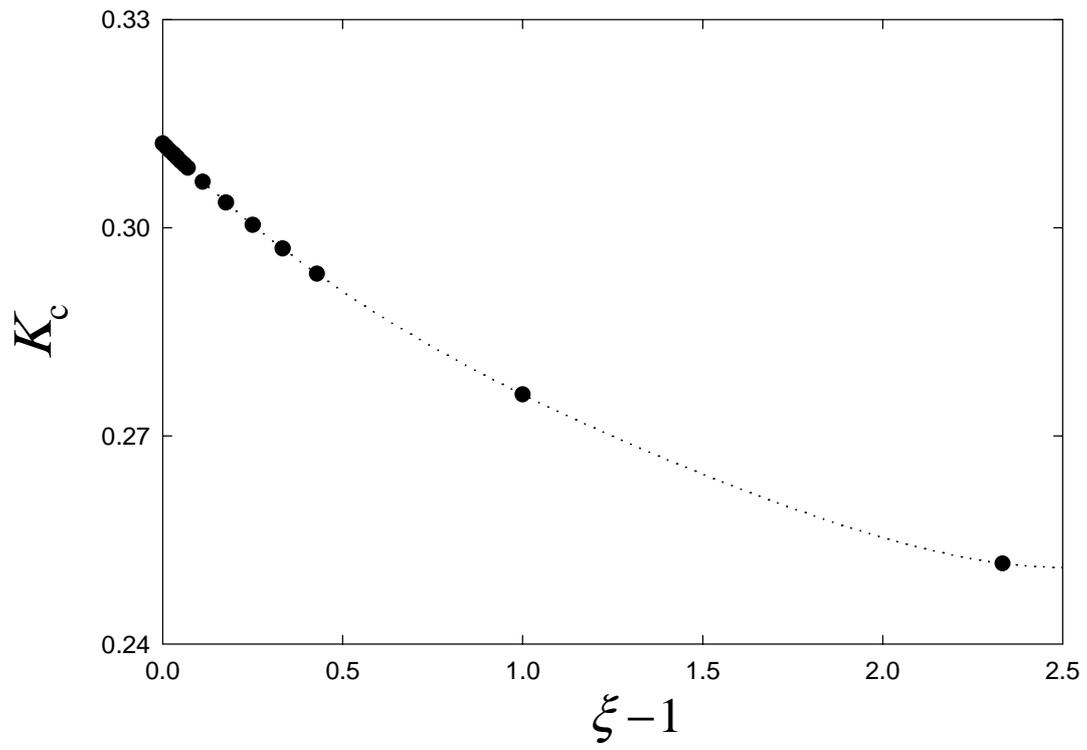